# Multi-Kernel TOF-PET Image Reconstruction Using ADMM

K. Ote, F. Hashimoto, Y. Onishi, R. Ota, M. Toussaint and F. Loignon-Houle

***Abstract*—Time-of-flight positron emission tomography (TOF-PET) detectors exhibiting multiple coincidence time resolution (CTR) components, such as those induced by the mixing of Cherenkov and scintillation photons, have attracted increasing attention. However, to fully exploit the latent potential of multi-kernel TOF-PET, new iterative image reconstruction methods are required. In this study, assuming that the events are labeled with the appropriate kernels, we propose an alternating direction method of multipliers (ADMM) for multi-kernel TOF PET reconstruction, termed TOF-decomp ADMM. As the convergence speed of the TOF-PET log-likelihood depends on the CTR, the proposed method splits the fast- and slow-CTR log-likelihood terms and optimizes them separately under a constraint. This strategy explicitly balances the contributions of fast- and slow-CTR components and enables early stopping at iterations that yield improved contrast–noise trade-offs compared with conventional methods. We validated the proposed method using brain and image quality phantom simulations, demonstrating improved contrast-noise characteristics from a more stabilized convergence. By addressing the convergence imbalance inherent to multi-kernel TOF-PET, this work establishes a framework for exploiting the timing information available in emerging detectors technologies.**

## I. Introduction

Time-of-flight positron emission tomography (TOF-PET) detectors that exhibit multiple coincidence time resolution (CTR) kernels, such as those induced by mixing of Cherenkov and scintillation photons [1], have attracted increasing attention in the PET research community. These detector concepts have been explored as candidates for cost-effective TOF-PET, notably bismuth germanate (BGO) crystals [2], as well as for enabling energy-sensitive measurements in direct positron emission imaging using scintillator-integrated microchannel plate photomultiplier tubes [3], [4]. Heterostructured scintillators, built on the principle of variable energy sharing between dense and fast materials in close proximity, also exhibit multiple CTR regimes [5]–[7].

These emerging multi-kernel TOF-PET detectors were studied theoretically to establish CTR limits, accounting for the variable amount of prompt photons produced event-by-event [8], [9]. Other investigations focused on signal-to-noise ratio (SNR) gain achieved by TOF information containing multiple CTRs [10], [11]. However, the analysis of TOF-related SNR gain becomes more complicated when image reconstruction is performed using iterative methods [12], compared with analytical approaches [13]. One of the reasons is because TOF information accelerates the convergence speed of iterative algorithms [14], which in turn leads to noise amplification as the number of iterations increases in maximum likelihood expectation maximization (MLEM) algorithms [15], [16]. Therefore, careful control of the number of iterations is required to achieve an optimal TOF gain in iterative TOF-PET image reconstruction.

This challenge becomes even more complex in multi-kernel TOF-PET image reconstruction, where events with different CTRs contribute different amounts of information and exhibit different convergence speeds [14]. To fully exploit the latent potential of multi-kernel TOF-PET detectors and systems, dedicated image reconstruction algorithms capable of stabilizing the convergence behavior across multiple CTR-dependent log-likelihood terms would be required. To address this, we propose a multi-kernel TOF-PET image reconstruction method using the alternating direction method of multipliers (ADMM) [17]. For simplicity, this study considers the case of two CTRs, comprising a fast CTR corresponding to coincidences in which both detections involve Cherenkov photons, and a slow CTR corresponding to coincidence in which both detections involve only scintillation photons. Here, we use the terms 'fast' and 'slow' loosely, with 'fast' denoting lower (better) CTR values and 'slow' denoting higher (worse) CTR values. A key advantage of the ADMM framework is that it allows the fast- and slow-CTR log-likelihood terms to be optimized separately while being softly coupled through a penalty term. This structure enables explicit and semi-quantitative control over the relative degree of optimization between the fast- and slow-CTR components, for example in principle by adjusting the number of sub-iterations assigned to each component or the step size, thereby providing

This work did not involve human subjects or animals in its research.

K. Ote, Y. Onishi, and R. Ota are with Central Research Laboratory, Hamamatsu Photonics K. K., 5000 Hirakuchi, Hamana-ku, Hamamatsu 434-8601, Japan. (e-mail: kibou@crl.hpk.co.jp, yuya.onishi@hpk.co.jp, ryosuke.ota@crl.hpk.co.jp).

F. Hashimoto is with J. Crayton Pruitt Family Department of Biomedical Engineering, University of Florida, Gainesville, Florida, USA. (e-mail: fumio.hashimo@ufl.edu).

M. Toussaint is with Laboratoire CRCI2NA, INSERM, CNRS, Nantes Université, Nantes, France. (e-mail: Maxime.Toussaint@inserm.fr).

F. Loignon-Houle is with Instituto de Instrumentación para Imagen Molecular (I3M), Consejo Superior de Investigaciones Científicas-Universitat Politècnica de València, Camino de Vera, Valencia 46002, Spain. (e-mail: floignon@i3m.upv.es).

a practical mechanism to balance their convergence behaviors.

We evaluated the proposed method using brain and image quality (IQ) phantom simulations. The results demonstrate that the proposed method facilitates early stopping at iterations associated with improved contrast–noise trade-offs, by stabilizing the convergence behavior between fast- and slow-CTR log-likelihoods.

## II. Backgrounds

### A. Measurement model of multi-kernel TOF-PET

The measurement model of multi-kernel TOF-PET is expressed as follows:

$$\begin{aligned} \boldsymbol{y} &\sim \mathrm{Poisson}(\bar{\boldsymbol{y}}), \\ \bar{\boldsymbol{y}} &= \boldsymbol{NA}\{\alpha \boldsymbol{P}_{\mathrm{fast}} + (1-\alpha)\boldsymbol{P}_{\mathrm{slow}}\}\boldsymbol{x}. \end{aligned} \tag{1}$$

Here, $\alpha$ denotes the expected fraction of events with fast CTR; $\boldsymbol{P}_{\mathrm{fast}} \in \mathbb{R}_+^{I\times J}$ and $\boldsymbol{P}_{\mathrm{slow}} \in \mathbb{R}_+^{I\times J}$ are projection matrices modeling fast and slow CTRs, respectively; $\boldsymbol{N} \in \mathbb{R}_+^{I\times I}$ and $\boldsymbol{A} \in \mathbb{R}_+^{I\times I}$ denote the diagonal matrices corresponding to normalization and attenuation factors, respectively; $\boldsymbol{y} \in \mathbb{Z}_+^{\boldsymbol{I}}$ and $\bar{\boldsymbol{y}} \in \mathbb{R}_+^{\boldsymbol{I}}$ are the measured and estimated sinograms, respectively; $\boldsymbol{x} \in \mathbb{R}_+^{\boldsymbol{J}}$ is the image to be reconstructed; and $I$ and $J$ are the total number of lines of response (LORs) and voxels, respectively. It is worth mentioning that, in the case of BGO, using a single global $\alpha$ shared across all LORs is sufficient, and that $\boldsymbol{P}_{\mathrm{fast}}$ and $\boldsymbol{P}_{\mathrm{slow}}$ differ only in terms of the TOF kernel. For simplicity, scatter and random coincidence events are not considered in this study.

If the events can be decomposed according to their CTRs, two measured sinograms are obtained:

$$\begin{aligned} \boldsymbol{y}_{\mathrm{fast}} &\sim \mathrm{Poisson}(\bar{\boldsymbol{y}}_{\mathrm{fast}}), & \bar{\boldsymbol{y}}_{\mathrm{fast}} &= \alpha \boldsymbol{NAP}_{\mathrm{fast}}\boldsymbol{x}, \\ \boldsymbol{y}_{\mathrm{slow}} &\sim \mathrm{Poisson}(\bar{\boldsymbol{y}}_{\mathrm{slow}}), & \bar{\boldsymbol{y}}_{\mathrm{slow}} &= (1-\alpha) \boldsymbol{NAP}_{\mathrm{slow}}\boldsymbol{x}. \end{aligned} \tag{2}$$

### B. TOF-mix MLEM

When the events cannot be decomposed according to their CTRs, we consider the following image reconstruction problem:

$$\max_{\boldsymbol{x}} L(\boldsymbol{y}|\boldsymbol{x}), \quad L(\boldsymbol{y}|\boldsymbol{x}) = \sum_{i=1}^{I} y_i \log \bar{y}_i - \bar{y}_i. \tag{3}$$

A MLEM [9] algorithm for solving the problem in (3) is given by

$$\boldsymbol{x}^{(k+1)} = \frac{\boldsymbol{x}^{(k)}}{\boldsymbol{\omega}} \boldsymbol{P}^T \frac{\boldsymbol{y}}{\boldsymbol{AP}\boldsymbol{x}^{(k)}}, \quad \boldsymbol{\omega} = \boldsymbol{P}^T \boldsymbol{N}\boldsymbol{1}, \tag{4}$$

where

$$\boldsymbol{P} = \alpha \boldsymbol{P}_{\mathrm{fast}} + (1-\alpha)\boldsymbol{P}_{\mathrm{slow}}. \tag{5}$$

Note that the attenuation factor $\boldsymbol{A}$ is applied only in the forward model as a practical implementation choice [18], in order to reuse the sensitivity image $\boldsymbol{\omega}$ [19]. The normalization factor $\boldsymbol{N}$ in the forward and backward projections is canceled out in the MLEM update and remains only in the sensitivity image. Accordingly, the sensitivity image is computed without the attenuation factor, as shown in (4). The algorithm in (4) is referred to as the **TOF-mix MLEM algorithm**.

### C. TOF-decomp MLEM

When the events can be decomposed according to their CTRs [1], we consider the following image reconstruction problem:

$$\begin{aligned} &\max_{\boldsymbol{x}} \{L_{\mathrm{fast}}(\boldsymbol{y}_{\mathrm{fast}}|\boldsymbol{x}) + L_{\mathrm{slow}}(\boldsymbol{y}_{\mathrm{slow}}|\boldsymbol{x})\}, \\ &L_{\mathrm{fast}}(\boldsymbol{y}_{\mathrm{fast}}|\boldsymbol{x}) = \sum_{i=1}^{I} y_{\mathrm{fast},i} \log \bar{y}_{\mathrm{fast},i} - \bar{y}_{\mathrm{fast},i}, \\ &L_{\mathrm{slow}}(\boldsymbol{y}_{\mathrm{slow}}|\boldsymbol{x}) = \sum_{i=1}^{I} y_{\mathrm{slow},i} \log \bar{y}_{\mathrm{slow},i} - \bar{y}_{\mathrm{slow},i}. \end{aligned} \tag{6}$$

A MLEM algorithm for solving the problem in (6) is given by

$$\boldsymbol{x}^{(k+1)} = \frac{\boldsymbol{x}^{(k)}}{\boldsymbol{\omega}} \left\{ \boldsymbol{P}_{\mathrm{fast}}^T \frac{\boldsymbol{y}_{\mathrm{fast}}}{\boldsymbol{AP}_{\mathrm{fast}}\boldsymbol{x}^{(k)}} + \boldsymbol{P}_{\mathrm{slow}}^T \frac{\boldsymbol{y}_{\mathrm{slow}}}{\boldsymbol{AP}_{\mathrm{slow}}\boldsymbol{x}^{(k)}} \right\}. \tag{7}$$

The algorithm in (7) is referred to as the **TOF-decomp MLEM algorithm**.

## III. Proposed Method

The MLEM algorithm exhibits increased noise as it approaches the maximum-likelihood solution. To mitigate this effect, early stopping is commonly employed. However, since the convergence speed of the MLEM algorithm depends on the CTR, selecting an optimal number of iterations that is suitable for both $L_{\mathrm{fast}}$ and $L_{\mathrm{slow}}$ could be challenging.

In this context, we propose an alternative optimization framework for $L_{\mathrm{fast}}$ and $L_{\mathrm{slow}}$ using the ADMM [17]. This framework enables control of the relative degree of convergence between $L_{\mathrm{fast}}$ and $L_{\mathrm{slow}}$, in principle, by adjusting the number of sub-iterations and step size. This algorithm is hereafter referred to as the **TOF-decomp ADMM algorithm**.

### A. TOF-decomp ADMM

We consider the following constrained optimization problem:

$$\max_{\boldsymbol{x},\boldsymbol{z}} \{L_{\mathrm{fast}}(\boldsymbol{y}_{\mathrm{fast}}|\boldsymbol{x}) + L_{\mathrm{slow}}(\boldsymbol{y}_{\mathrm{slow}}|\boldsymbol{z})\} \quad \mathrm{s.t.} \quad \boldsymbol{x} = \boldsymbol{z}. \tag{8}$$

Using the augmented Lagrangian, the constrained optimization problem is transformed into the following unconstrained problem:

$$\max_{x,z}\begin{Bmatrix}L_{\text{fast}}(\boldsymbol{y}_{\text{fast}}|\boldsymbol{x})+L_{\text{slow}}(\boldsymbol{y}_{\text{slow}}|\boldsymbol{z})\\ -\frac{1}{\rho}\boldsymbol{\mu}^T(\boldsymbol{x}-\boldsymbol{z})-\frac{1}{2\rho}\|\boldsymbol{x}-\boldsymbol{z}\|^2\end{Bmatrix}. \tag{9}$$

where $\boldsymbol{\mu}\in\mathbb{R}^J$ denotes the scaled Lagrange multiplier and $\rho$ is a positive constant corresponding to the step size. By completing the square, we obtain:

$$\begin{aligned}(\boldsymbol{x}^*,\boldsymbol{z}^*)&=\underset{x,z}{\operatorname{argmax}}\begin{Bmatrix}L_{\text{fast}}(\boldsymbol{y}_{\text{fast}}|\boldsymbol{x})+L_{\text{slow}}(\boldsymbol{y}_{\text{slow}}|\boldsymbol{z})\\ -\frac{1}{2\rho}\|\boldsymbol{x}-\boldsymbol{z}+\boldsymbol{\mu}\|^2+\frac{1}{2\rho}\|\boldsymbol{\mu}\|^2\end{Bmatrix}\\ &=\underset{x,z}{\operatorname{argmax}}\begin{Bmatrix}L_{\text{fast}}(\boldsymbol{y}_{\text{fast}}|\boldsymbol{x})+L_{\text{slow}}(\boldsymbol{y}_{\text{slow}}|\boldsymbol{z})\\ -\frac{1}{2\rho}\|\boldsymbol{x}-\boldsymbol{z}+\boldsymbol{\mu}\|^2\end{Bmatrix}.\end{aligned} \tag{10}$$

Here, second line of Eq. (10) is obtained from the fact that the term $\frac{1}{2\rho}\|\boldsymbol{\mu}\|^2$ is not relevant to the parameters $(\boldsymbol{x},\boldsymbol{z})$.

ADMM solves the problem (10) by iteratively performing the following three steps:

$$\boldsymbol{z}^{(k+1)}=\underset{z}{\operatorname{argmax}}\left\{L_{\text{slow}}(\boldsymbol{y}_{\text{slow}}|\boldsymbol{z})-\frac{1}{2\rho}\left\|\begin{matrix}\boldsymbol{x}^{(k)}-\boldsymbol{z}\\+\boldsymbol{\mu}^{(k)}\end{matrix}\right\|^2\right\}, \tag{11}$$

$$\boldsymbol{x}^{(k+1)}=\underset{x}{\operatorname{argmax}}\left\{L_{\text{fast}}(\boldsymbol{y}_{\text{fast}}|\boldsymbol{x})-\frac{1}{2\rho}\left\|\begin{matrix}\boldsymbol{x}-\boldsymbol{z}^{(k+1)}\\+\boldsymbol{\mu}^{(k)}\end{matrix}\right\|^2\right\}, \tag{12}$$

$$\boldsymbol{\mu}^{(k+1)}=\boldsymbol{\mu}^{(k)}+\boldsymbol{x}^{(k+1)}-\boldsymbol{z}^{(k+1)}. \tag{13}$$

When solving subproblems (11) and (12), different numbers of sub-iterations can be chosen to balance the degree of optimization for $L_{\text{slow}}$ and $L_{\text{fast}}$. This flexibility constitutes the main advantage of the proposed algorithm.

### *B. Solving subproblems*

The subproblems (11) and (12) can be solved by minorize-maximization (MM) algorithm [20]. In the MM framework, $L_{\text{slow}}$ and $L_{\text{fast}}$ are replaced by the following surrogate functions [21]:

$$\begin{aligned}Q_{\text{slow}}\left(\boldsymbol{z}\middle|\boldsymbol{z}^{(k)}\right)&=\sum_{j=1}^{J}\omega_{\text{slow},j}\left(z_{\text{EM},j}^{(k+1)}\log z_j-z_j\right),\\ Q_{\text{fast}}\left(\boldsymbol{x}\middle|\boldsymbol{x}^{(k)}\right)&=\sum_{j=1}^{J}\omega_{\text{fast},j}\left(x_{\text{EM},j}^{(k+1)}\log x_j-x_j\right),\end{aligned} \tag{14}$$

Here,

$$\begin{aligned}\boldsymbol{z}_{\text{EM}}^{(k+1)}&=\frac{\boldsymbol{z}^{(k)}}{\boldsymbol{\omega}_{\text{slow}}}\boldsymbol{P}_{\text{slow}}^T\frac{\boldsymbol{y}_{\text{slow}}}{\boldsymbol{A}\boldsymbol{P}_{\text{slow}}\boldsymbol{z}^{(k)}},\\ \boldsymbol{x}_{\text{EM}}^{(k+1)}&=\frac{\boldsymbol{x}^{(k)}}{\boldsymbol{\omega}_{\text{fast}}}\boldsymbol{P}_{\text{fast}}^T\frac{\boldsymbol{y}_{\text{fast}}}{\boldsymbol{A}\boldsymbol{P}_{\text{fast}}\boldsymbol{x}^{(k)}},\\ \boldsymbol{\omega}_{\text{slow}}&=(1-\alpha)\boldsymbol{P}_{\text{slow}}^T\boldsymbol{N}\boldsymbol{1},\quad \boldsymbol{\omega}_{\text{fast}}=\alpha\boldsymbol{P}_{\text{fast}}^T\boldsymbol{N}\boldsymbol{1}.\end{aligned} \tag{15}$$

The surrogate functions $Q_{\text{slow}}$ and $Q_{\text{fast}}$ satisfy the following conditions:

$$\begin{aligned}&\nabla L_{\text{slow}}\left(\boldsymbol{y}_{\text{slow}}\middle|\boldsymbol{z}^{(k)}\right)=\nabla Q_{\text{slow}}\left(\boldsymbol{z}^{(k)}\middle|\boldsymbol{z}^{(k)}\right),\\ &\begin{Bmatrix}L_{\text{slow}}(\boldsymbol{y}_{\text{slow}}|\boldsymbol{x})-\\ L_{\text{slow}}\left(\boldsymbol{y}_{\text{slow}}\middle|\boldsymbol{z}^{(k)}\right)\end{Bmatrix}\geq\begin{Bmatrix}Q_{\text{slow}}\left(\boldsymbol{z}\middle|\boldsymbol{z}^{(k)}\right)-\\ Q_{\text{slow}}\left(\boldsymbol{z}^{(k)}\middle|\boldsymbol{z}^{(k)}\right)\end{Bmatrix},\\ &\nabla L_{\text{fast}}\left(\boldsymbol{y}_{\text{fast}}\middle|\boldsymbol{x}^{(k)}\right)=\nabla Q_{\text{fast}}\left(\boldsymbol{x}^{(k)}\middle|\boldsymbol{x}^{(k)}\right),\\ &\begin{Bmatrix}L_{\text{fast}}(\boldsymbol{y}_{\text{fast}}|\boldsymbol{x})-\\ L_{\text{fast}}\left(\boldsymbol{y}_{\text{fast}}\middle|\boldsymbol{x}^{(k)}\right)\end{Bmatrix}\geq\begin{Bmatrix}Q_{\text{fast}}\left(\boldsymbol{x}\middle|\boldsymbol{x}^{(k)}\right)-\\ Q_{\text{fast}}\left(\boldsymbol{x}^{(k)}\middle|\boldsymbol{x}^{(k)}\right)\end{Bmatrix}.\end{aligned} \tag{16}$$

Therefore, maximizing $Q_{\text{slow}}$ and $Q_{\text{fast}}$ guarantees a monotonic increase in $L_{\text{slow}}$ and $L_{\text{fast}}$, respectively. By replacing $L_{\text{slow}}$ and $L_{\text{fast}}$ with $Q_{\text{slow}}$ and $Q_{\text{fast}}$, the objective functions become separable voxel by voxel:

$$\begin{aligned}&\omega_{\text{slow},j}\left(z_{\text{EM},j}^{(k+1)}\log z_j-z_j\right)-\frac{1}{2\rho}\left(x_j^{(k)}-z_j+\mu_j^{(k)}\right)^2,\\ &\omega_{\text{fast},j}\left(x_{\text{EM},j}^{(k+1)}\log x_j-x_j\right)-\frac{1}{2\rho}\left(x_j-z_j^{(k+1)}+\mu_j^{(k)}\right)^2.\end{aligned} \tag{17}$$

By setting the derivatives of (17) to zero, closed-form solutions are obtained:

$$\begin{aligned}z_j^{(k+1)}&=\frac{b_j^{(k)}+\sqrt{\left(b_j^{(k)}\right)^2+4z_{\text{EM},j}^{(k+1)}\rho\omega_{\text{slow},j}}}{2},\\ b_j^{(k)}&=x_j^{(k)}+\mu_j^{(k)}-\rho\omega_{\text{slow},j},\\ x_j^{(k+1)}&=\frac{c_j^{(k)}+\sqrt{\left(c_j^{(k)}\right)^2+4x_{\text{EM},j}^{(k+1)}\rho\omega_{\text{fast},j}}}{2},\\ c_j^{(k)}&=z_j^{(k+1)}-\mu_j^{(k)}-\rho\omega_{\text{fast},j}.\end{aligned} \tag{18}$$

Since the variable is constrained to be nonnegative, we retain only the positive root of the quadratic equation. Thanks to the elegance of the ADMM and MM frameworks, we thus derive a new algorithm for multi-kernel TOF-PET image reconstruction (**Algorithm 1**).

## IV. Simulation Setup

We evaluated the proposed method using Monte Carlo simulation data of a brain TOF-PET scanner (Hamamatsu HIAS-29000) [22]. The scanner consists of 28 units, each comprising four detector modules. Each module consists of 16 × 16 lutetium fine silicate (LFS) crystals with dimensions of 3.14 × 3.14 × 20 mm³, coupled to 16 × 16 multi-pixel photon counters (MPPCs) as shown in **Fig. 1**. In the simulation, we initially assumed a CTR of 0 ps. Gaussian noise is then added to the TOF information depending on whether the event has a fast or slow CTR. In this study, we assume fast and slow CTRs of 70 ps and 400 ps in full width at half maximum, respectively (**Fig. 2**), adopting the values employed in [10], a prior study on SNR gain from multi-TOF kernels. We considered four cases of fraction of events with fast CTR with $\alpha=0.1, 0.25, 0.5,$ and $0.75$.

**Algorithm 1 TOF-decomp ADMM**

1: $\boldsymbol{x}^{(0)}, \boldsymbol{z}^{(0)} \in \mathbb{R}_+^J, \boldsymbol{\mu}^{(0)} = \boldsymbol{0}$.
2: **for** $k = 0$ **to** $N-1$ **do**

Solving subproblem (11)

3: $\boldsymbol{z}^{(k,0)} = \boldsymbol{z}^{(k)}, \quad \boldsymbol{b}^{(k)} = \boldsymbol{x}^{(k)} + \boldsymbol{\mu}^{(k)} - \rho\boldsymbol{\omega}_{\text{slow}}$.
4: **for** $m = 0$ **to** $M_1 - 1$ **do**
5: $\boldsymbol{z}_{\text{EM}}^{(k,m+1)} = \frac{\boldsymbol{z}^{(k,m)}}{\boldsymbol{\omega}_{\text{slow}}} \boldsymbol{P}_{\text{slow}}^T \frac{\boldsymbol{y}_{\text{slow}}}{\boldsymbol{A}\boldsymbol{P}_{\text{slow}}\boldsymbol{z}^{(k)}}$.
6: $z_j^{(k,m+1)} = \frac{b_j^{(k)} + \sqrt{\left(b_j^{(k)}\right)^2 + 4z_{\text{EM},j}^{(k,m+1)}\rho\omega_{\text{slow},j}}}{2}$.
7: **end for**
8: $\boldsymbol{z}^{(k+1)} = \boldsymbol{z}^{(k,M_1)}$.

Solving subproblem (12)

9: $\boldsymbol{x}^{(k,0)} = \boldsymbol{x}^{(k)}, \quad \boldsymbol{c}^{(k)} = \boldsymbol{z}^{(k+1)} - \boldsymbol{\mu}^{(k)} - \rho\boldsymbol{\omega}_{\text{fast}}$.
10: **for** $m = 0$ **to** $M_2 - 1$ **do**
11: $\boldsymbol{x}_{\text{EM}}^{(k,m+1)} = \frac{\boldsymbol{x}^{(k)}}{\boldsymbol{\omega}_{\text{fast}}} \boldsymbol{P}_{\text{fast}}^T \frac{\boldsymbol{y}_{\text{fast}}}{\boldsymbol{A}\boldsymbol{P}_{\text{fast}}\boldsymbol{x}^{(k,m)}}$.
12: $x_j^{(k,m+1)} = \frac{c_j^{(k)} + \sqrt{\left(c_j^{(k)}\right)^2 + 4x_{\text{EM},j}^{(k,m+1)}\rho\omega_{\text{fast},j}}}{2}$.
13: **end for**
14: $\boldsymbol{x}^{(k+1)} = \boldsymbol{x}^{(k,M_2)}$.

Update Lagrange multiplier

15: $\boldsymbol{\mu}^{(k+1)} = \boldsymbol{\mu}^{(k)} + \boldsymbol{x}^{(k+1)} - \boldsymbol{z}^{(k+1)}$.
16: **end for**

*A. Brain phantom*

A digital brain phantom was generated from a segmented magnetic resonance imaging (MRI) data downloaded from BrainWeb [23]. The activity contrast was set to 1:0.25:0.05 for gray matter, white matter, and cerebrospinal fluid, respectively. Three hot spots mimicking tumors were embedded with activity contrasts of 1.5, 1.2, and 1.1, and radii of 1.0, 1.2, and 1.6 cm, respectively. The attenuation coefficients were set to 0.00958 mm$^{-1}$ and 0.0151 mm$^{-1}$ for soft tissue and bone, respectively.

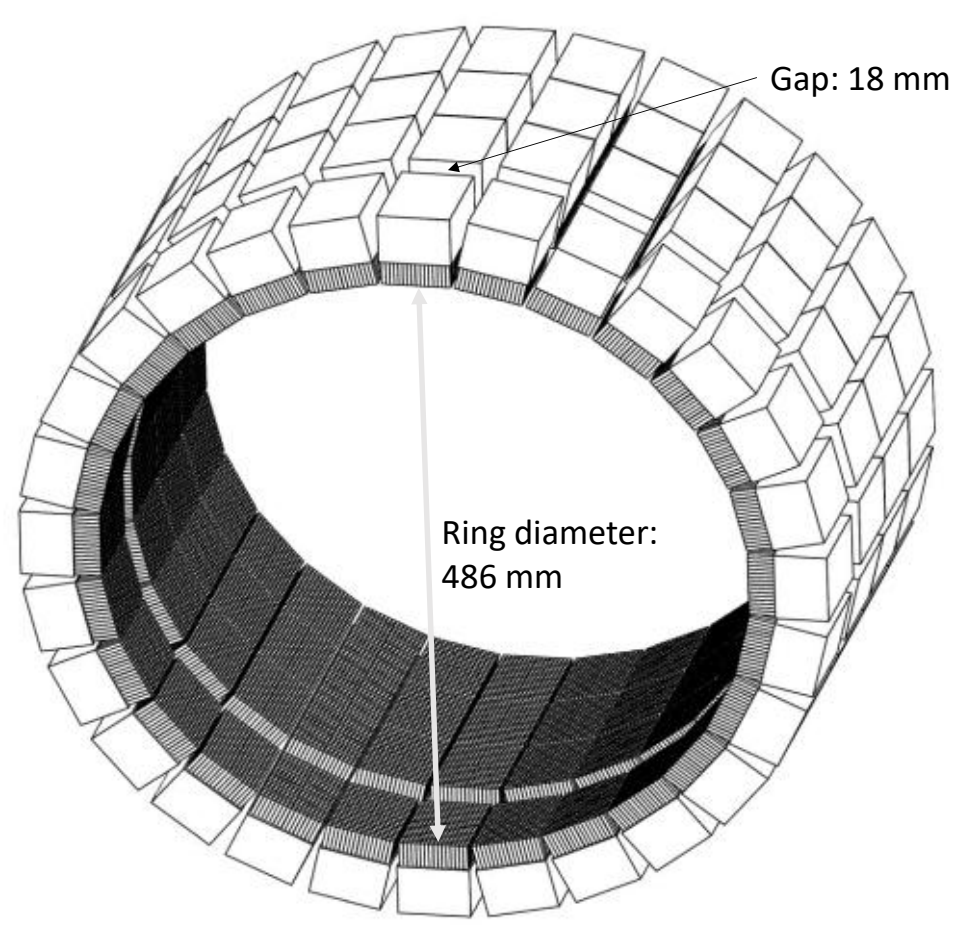

**Fig. 1.** Mechanical drawing illustrating the detector arrangement of the HIAS-29000 brain PET scanner.

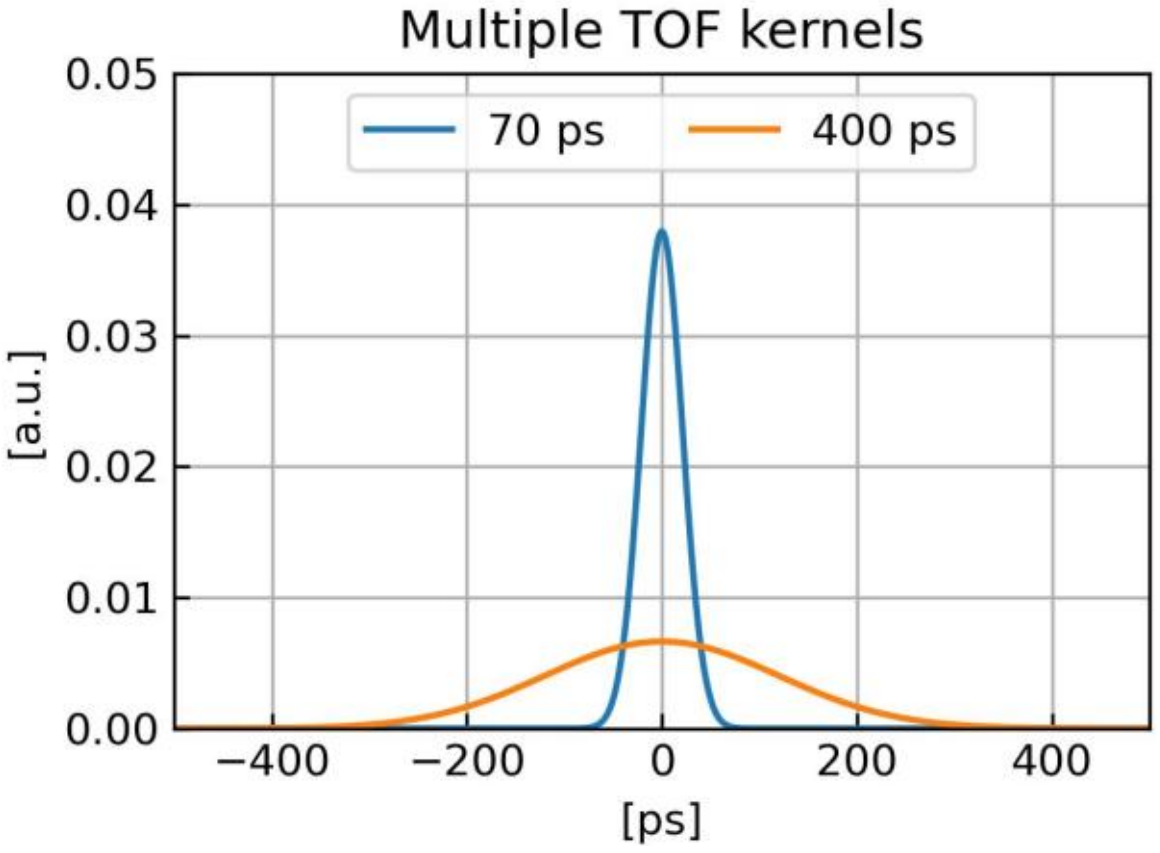

**Fig. 2.** Multiple TOF kernels in this study.

For simplicity, events involving Compton scattering were rejected, resulting in a total of 117 million events. These events were divided into 10 non-overlapping subsets to generate 10 samples, each containing 11.7 million events.

*B. Image quality phantom*

A uniform cylinder containing six spheres, as in the NEMA image quality (IQ) phantom [24], was simulated. The diameter of the cylinder was 20 cm, and the contrast between hot spheres and background was set to 4:1. The length of cylinder was 23.03 cm. The attenuation coefficient inside the cylinder was set to 0.00958 mm$^{-1}$. As in the brain phantom, scattered events were also rejected, resulting in 98.4 million events divided into 10 samples of 9.8 million events each.

*C. Implementation details*

In this study, we implemented TOF-mix/decomp MLEM and TOF-decomp ADMM in list-mode PET image reconstruction [25], [26] using parallelproj [27]. We ran all algorithms until 50 iterations. We set $M_1 = 2$, $M_2 = 1$, and $\rho = 0.02$ for TOF-decomp ADMM (**Algorithm 1**) to balance the convergence speeds between 70 and 400 ps CTR components and to stabilize the alternating updates against statistical noise corresponding to 11.7 and 9.8 million events. For comparison, we also evaluated the TOF-decomp ADMM with a symmetric update of $M_1 = M_2 = 1$, denoted as **TOF-decomp ADMM-sym**. Image and voxel sizes are 128 × 128 × 70 voxels and 3.0 × 3.0 × 3.221 mm$^3$, respectively.

*D. Evaluation metrics*

Peak signal to noise ratio (PSNR) was evaluated for both brain and IQ phantoms:

$$\text{PSNR} = 20\log_{10} \frac{\max(\boldsymbol{x}_{\text{pha}})}{\sqrt{\frac{1}{N_{\text{in}}}\sum_{j\in\text{ROI}_{\text{in}}}\left(x_{\text{pha},j} - x_j\right)^2}}, \tag{19}$$

where $\boldsymbol{x}_{\text{pha}}$ is a phantom image, $\text{ROI}_{\text{in}}$ is a region of interest (ROI) inside the phantom, and $N_{\text{in}}$ is number of voxels inside the phantom.

The tumor uptake ratio (TR) was evaluated for the brain phantom:

$$\mathrm{TR} = \frac{\sum_{n=1}^{3}\sum_{j\in\mathrm{ROI}_{\mathrm{tum},n}} x_j}{\sum_{n=1}^{3}\sum_{j\in\mathrm{ROI}_{\mathrm{tum},n}} x_{\mathrm{pha},j}} \times 100\%, \tag{20}$$

where $\mathrm{ROI}_{\mathrm{tum},n}$ denotes the ROI on the $n$-th tumor.

The contrast recovery coefficient (CRC) was evaluated for each hot sphere in the IQ phantom [28]:

$$\mathrm{CRC}_{\mathrm{hot}} = \frac{\mathrm{mean}_{\mathrm{hot}}}{\frac{1}{6}\sum_{n=1}^{6}\mathrm{mean}_{\mathrm{bg},n}}\frac{1}{4} \times 100\%, \tag{21}$$

where $\mathrm{mean}_{\mathrm{hot}}$ is the mean value within the ROI on a hot sphere, and $\mathrm{mean}_{\mathrm{bg},n}$ is a mean value within the ROI of the $n$-th background region. Six background ROIs with a diameter of 2 cm were placed near the six spheres in the same slice.

The CRC was also evaluated for each cold sphere in the IQ phantom [28]:

$$\mathrm{CRC}_{\mathrm{cold}} = 1 - \frac{\mathrm{mean}_{\mathrm{cold}}}{\frac{1}{6}\sum_{n=1}^{6}\mathrm{mean}_{\mathrm{bg,n}}} \times 100\%, \tag{22}$$

where $\mathrm{mean}_{\mathrm{cold}}$ is the mean value within the ROI on a cold sphere.

The coefficient of variation (COV) was evaluated for the IQ phantom at each slice:

$$\mathrm{COV}_{\mathrm{slice}} = \frac{\mathrm{StdDev}_{\mathrm{slice}}}{\mathrm{Mean}_{\mathrm{slice}}} \times 100\%, \tag{23}$$

where $\mathrm{Mean}_{\mathrm{slice}}$ and $\mathrm{StdDev}_{\mathrm{slice}}$ are mean and standard deviation, respectively, calculated in the central circular region with a diameter of 6 cm.

Sample mean and standard deviation images were calculated from 10 samples of simulation data:

$$\bar{\boldsymbol{x}} = \frac{1}{R}\sum_{r=1}^{R}\boldsymbol{x}^r, \quad s_j = \sqrt{\frac{1}{R}\sum_{r=1}^{R}\left(x_j^r - \bar{x}_j\right)^2}, \tag{24}$$

where $\boldsymbol{x}^r$ is a reconstructed image from $r$-th sample of simulation data, and $R = 10$ in this study.

Sample bias and variance were evaluated for both brain and IQ phantoms:

$$\begin{aligned}\mathrm{Bias}^2 &= \frac{1}{N_{\mathrm{in}}}\sum_{j\in\mathrm{ROI}_{\mathrm{in}}}\left(x_{\mathrm{pha},j} - \bar{x}_j\right)^2, \\ \mathrm{Var} &= \frac{1}{N_{\mathrm{in}}}\sum_{j\in\mathrm{ROI}_{\mathrm{in}}} s_j^2 .\end{aligned} \tag{25}$$

In addition, the root mean squared error (RMSE) was evaluated for the brain phantom:

$$\mathrm{RMSE} = \sqrt{\mathrm{Bias}^2 + \mathrm{Var}}. \tag{26}$$

To visualize the convergence behavior, the progress of the fast- and slow-CTR log-likelihoods was evaluated as:

$$\begin{aligned} p_{\mathrm{fast}}^{(k)} &= L_{\mathrm{fast}}\left(U_{\mathrm{fast}}\middle|\boldsymbol{x}^{(k+1)}\right) - L_{\mathrm{fast}}\left(U_{\mathrm{fast}}\middle|\boldsymbol{x}^{(k)}\right), \\ p_{\mathrm{slow}}^{(k)} &= L_{\mathrm{slow}}\left(U_{\mathrm{slow}}\middle|\boldsymbol{x}^{(k+1)}\right) - L_{\mathrm{slow}}\left(U_{\mathrm{fast}}\middle|\boldsymbol{x}^{(k)}\right), \end{aligned} \tag{27}$$

where $U_{\mathrm{fast}}$ and $U_{\mathrm{slow}}$ are the list-mode data of fast- and slow-CTR events, respectively, and $L_{\mathrm{fast}}(U_{\mathrm{fast}}|\boldsymbol{x})$ and $L_{\mathrm{slow}}(U_{\mathrm{slow}}|\boldsymbol{x})$ denotes the corresponding list-mode log-likelihoods.

The relative progress was evaluated as:

$$r^{(k)} = \frac{p_{\mathrm{fast}}^{(k)}}{p_{\mathrm{slow}}^{(k)}}. \tag{28}$$

## V. Results

**Fig. 3** presents the PSNR-TR and $\mathrm{Bias}^2$-Var trade-off curves, as well as RMSE curves, obtained from the brain phantom simulations, where PSNR and TR values are averaged over 10 samples. The TOF-decomp ADMM method consistently achieves more favorable trade-offs than both the TOF-mix MLEM and TOF-decomp MLEM methods across various fast-CTR event ratios, indicating an improved balance between image quality metrics and robustness with respect to fast-CTR event ratio. In addition, most trade-off curves exhibit a clear saturation trend within the iteration range considered in this study (iterations 5–50), which is consistent with the accelerated convergence induced by TOF information. The RMSE curves show trends consistent with those observed in the $\mathrm{Bias}^2$-Var trade-off, further supporting the overall performance characteristics of each method. Notably, TOF-decomp ADMM exhibits flatter and wider RMSE minima compared with the other methods, suggesting a more stable optimal solution.

Representative brain images reconstructed at the iterations minimizing RMSE are shown in **Fig. 4** for $\alpha = 0.1$. This selection enables a fair visual comparison among the different methods and illustrates the practical impact of the improved trade-off observed in **Fig. 3.** While quantitative differences can be observed, the visual differences between the reconstruction methods remain relatively subtle.

The PSNR–CRC trade-off curves obtained from the IQ phantom simulations are shown in **Fig. 5** for $\alpha = 0.1$ and 0.5. For smaller hot spheres, TOF-decomp ADMM provides trade-off curves comparable to those of the other methods. Notably, for larger cold spheres (28 mm and 37 mm in diameter), the TOF-decomp MLEM and TOF-decomp ADMM methods achieve higher CRC values than the TOF-mix MLEM method at comparable PSNR levels, demonstrating that the classification of events into fast and slow CTR components is benefical for improving cold-sphere contrast recovery.

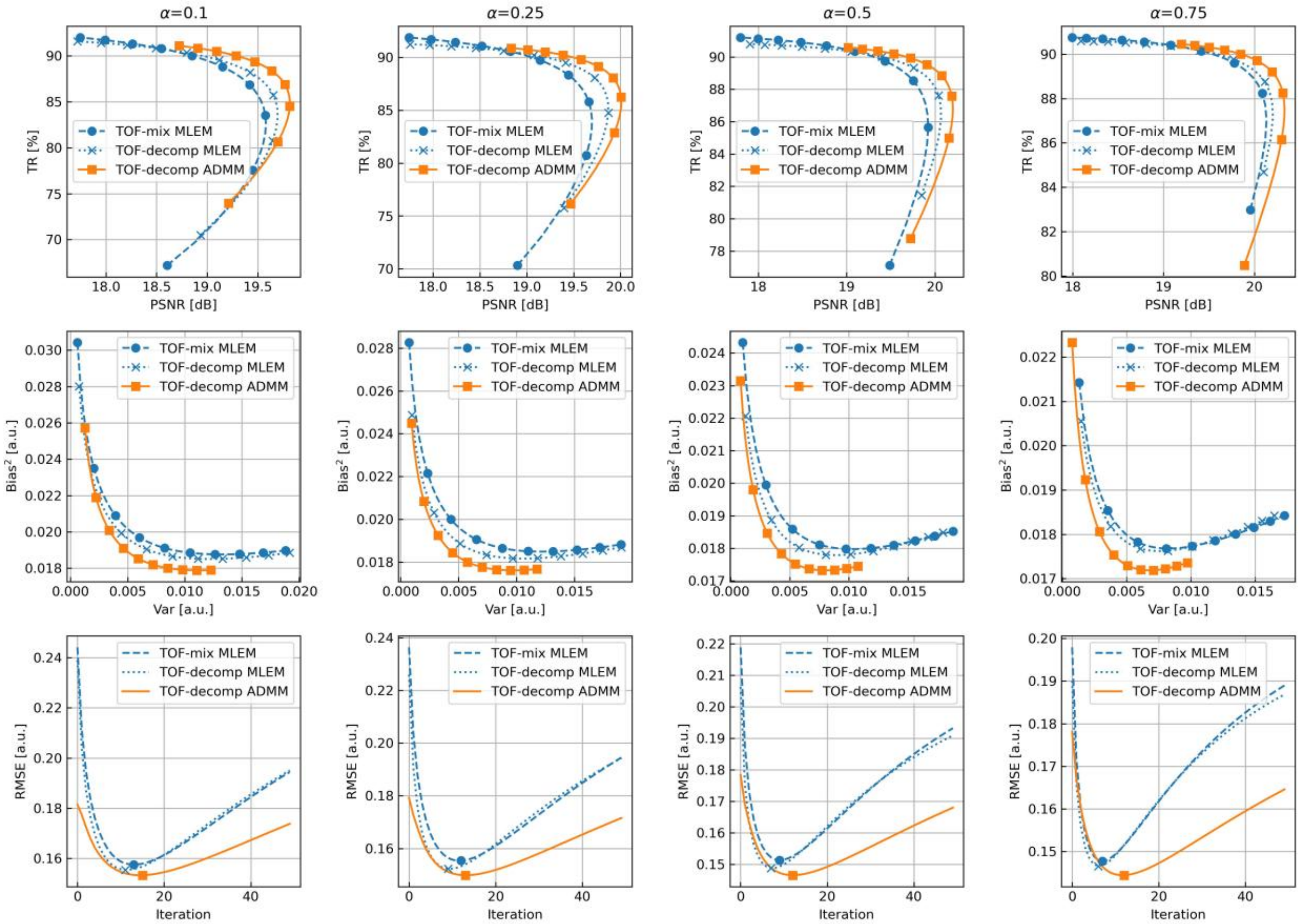

**Fig. 3.** PSNR-TR and Bias$^2$-Var trade-off curves, as well as RMSE curves, obtained from the brain simulation data for the four different fractions of events with fast CTR. PSNR and TR values are averaged over 10 samples. The markers correspond to every 5 iterations. The markers on the RMSE curves correspond to the images shown in **Fig. 4**.

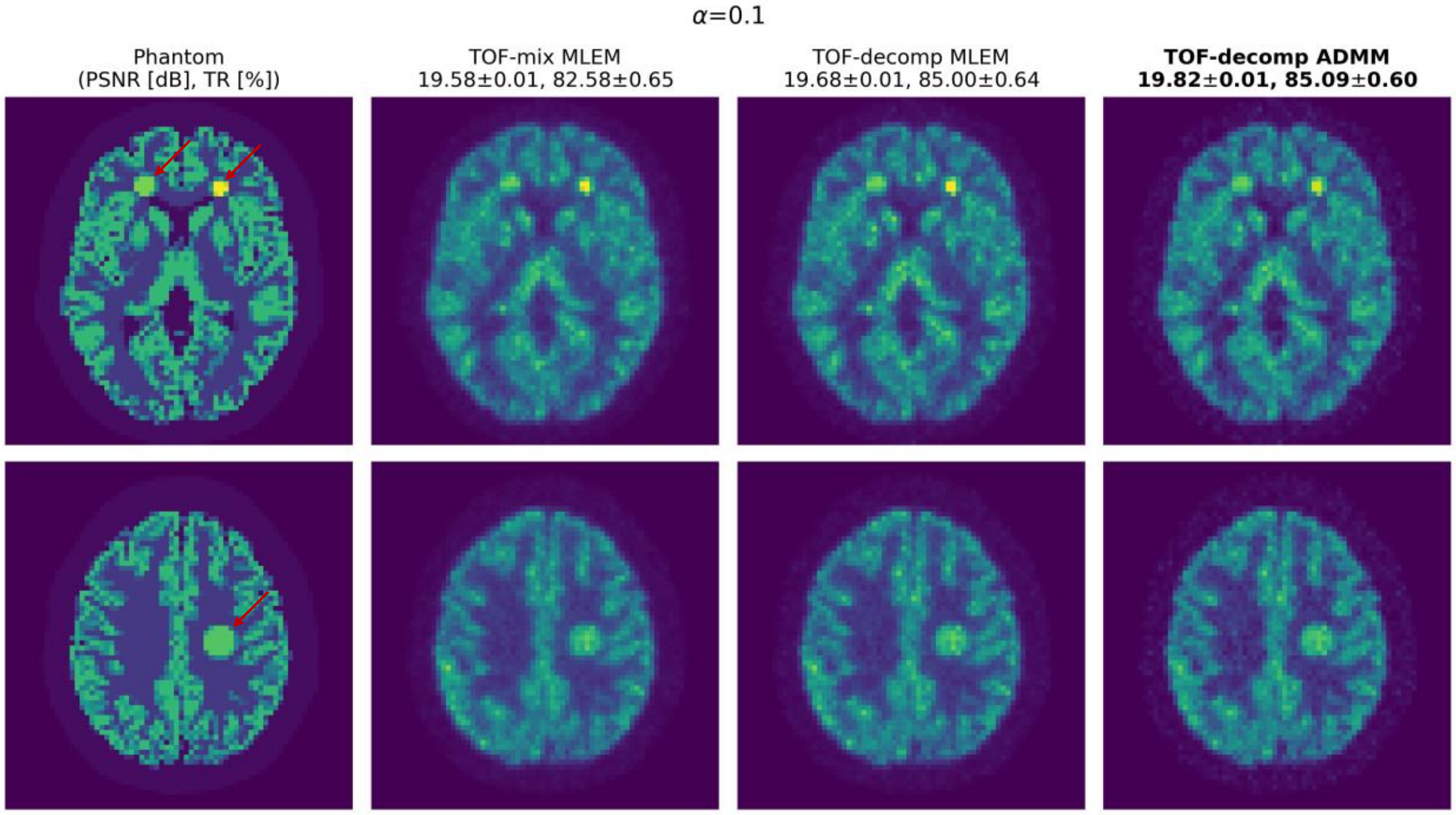

**Fig. 4.** Brain simulation images for $\alpha = 0.1$, shown at the iterations that achieve the minimum value of RMSE. The arrows indicate the hot spots mimicking tumors.

Furthermore, TOF-decomp ADMM tends to provide higher cold contrast recovery than TOF-decomp MLEM at similar PSNR levels, suggesting that the ADMM-based reconstruction may further enhance the trade-off between cold-sphere contrast

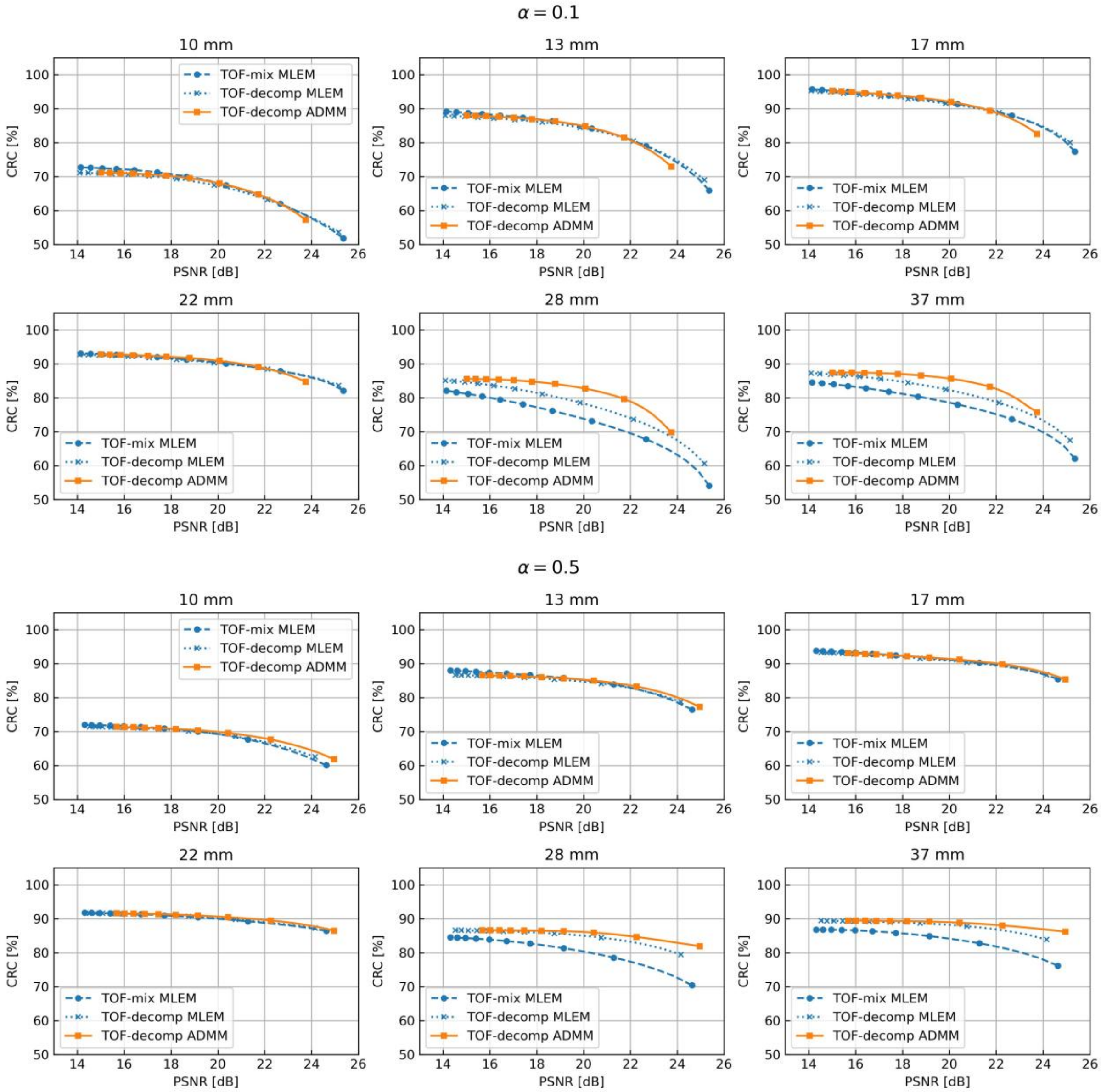

**Fig. 5.** PSNR-CRC trade-off curves obtained from the IQ simulation data for $\alpha = 0.1$ and 0.5 and the six different sphere diameters. PSNR and CRC values are averaged over 10 samples. The markers correspond to every 5 iterations.

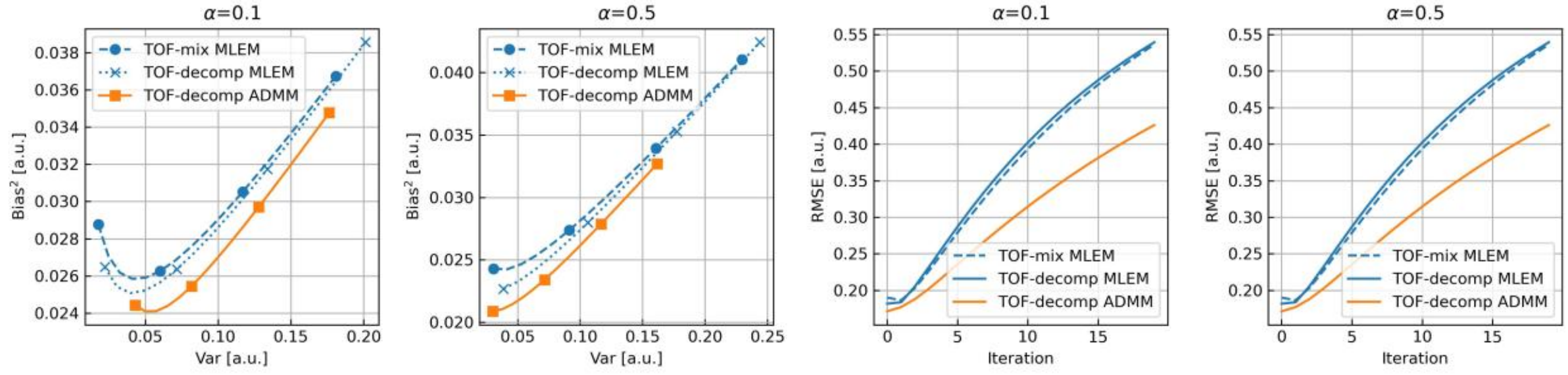

**Fig. 6.** Bias$^2$-Var trade-off curves and RMSE curves obtained from the IQ simulation data for $\alpha = 0.1$ and 0.5 up to 20 iterations. Markers indicate every 5 iterations.

recovery and background noise characteristics within the uniform cylindrical region**.**

**Figure 6** shows the Bias$^2$-Var trade-off curves and RMSE curves obtained from IQ simulation data for $\alpha = 0.1$ and 0.5. Compared with those obtained from the brain phantom simulations, these trade-off curves increase more rapidly with increasing iteration number. This behavior is attributed to the higher noise sensitivity of the IQ phantom, which has a simpler structure and a larger uniform background region than the brain phantom. Nevertheless, the TOF-decomp ADMM method still

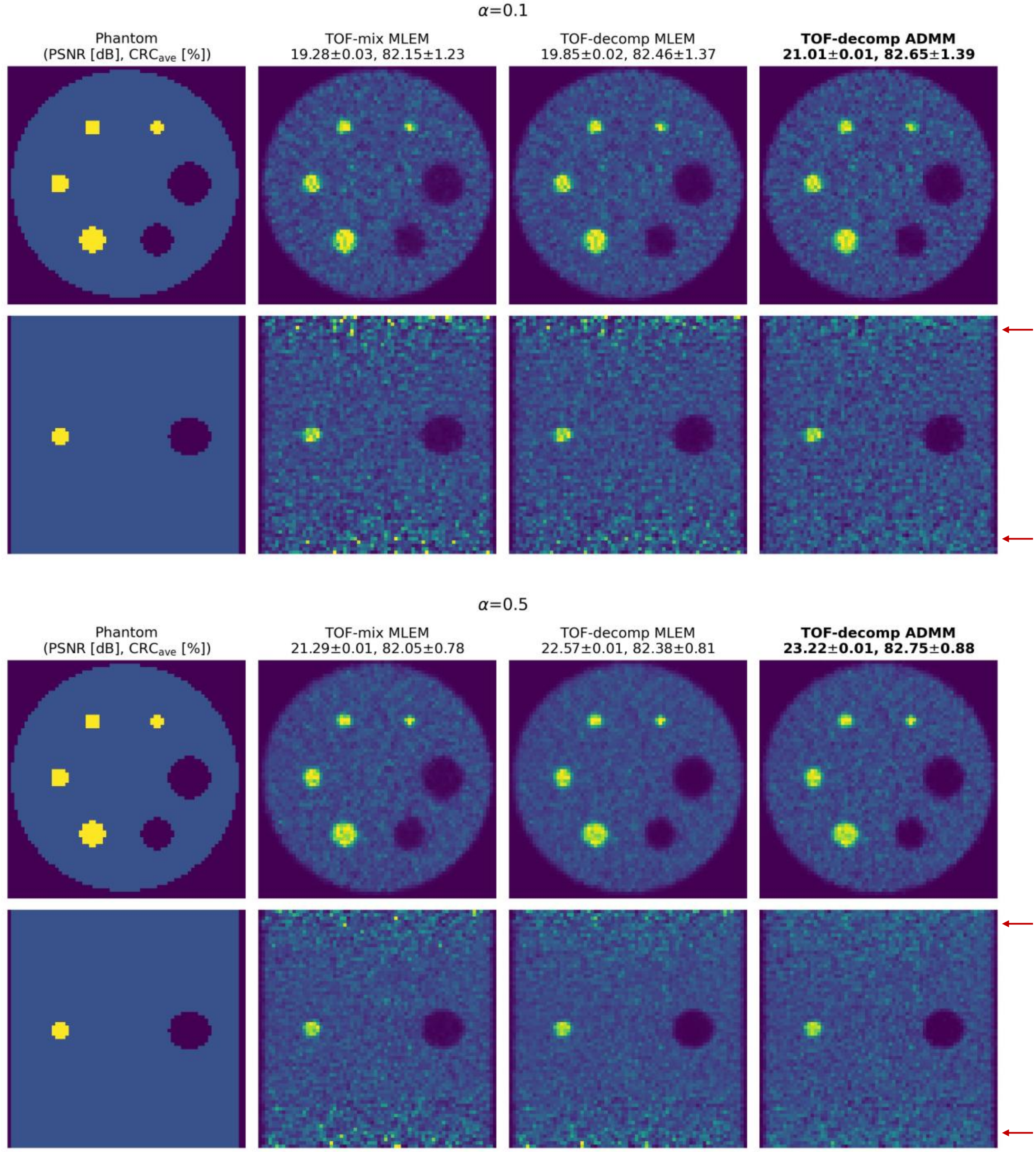

**Fig. 7.** IQ simulation images in transaxial and coronal planes for $\alpha = 0.1$ and $0.5$, shown at the iterations where $\mathrm{CRC_{ave}}$ reaches 95% of its maximum value, where $\mathrm{CRC_{ave}}$ denotes an average CRC over the six spheres. The TOF-decomp ADMM method exhibits visibly reduced noise in the axially peripheral regions in the coronal views, as indicated by the red arrows.

yields more favorable trade-offs than the TOF-mix MLEM and TOF-decomp MLEM methods.

Reconstructed IQ phantom images in transaxial and coronal planes are shown in **Fig. 7** for $\alpha = 0.1$ and $0.5$, at the iterations where $\mathrm{CRC_{ave}}$ reaches 95% of its maximum value, where $\mathrm{CRC_{ave}}$ denotes the average CRC over six spheres. We selected $\mathrm{CRC_{ave}}$ as the stopping criterion for the IQ simulation because the RMSE exhibits a sharp minimum at very early iterations, as shown in **Fig. 6**. At this stage, the contrast recovery and spatial resolution are still insufficient, resulting in slightly blurred image, which makes RMSE-based early stopping suboptimal for visual assessment. Notably, at the iteration where $\mathrm{CRC_{ave}}$ reaches 95% of its maximum value, the TOF-decomp ADMM method exhibits visibly reduced noise in the axially peripheral regions in the coronal views, as indicated by the red arrows.

To further quantify the noise, **Fig. 8** presents the slice-wise COV, evaluated at the iterations where $\mathrm{CRC_{ave}}$ reaches 95% of its maximum value. The TOF-decomp ADMM method shows a more uniform COV across slices compared with the TOF-mix MLEM and TOF-decomp MLEM methods, indicating improved noise consistency along the axial direction. The dip in the slice-wise COV, indicated by the red arrows, is attributed to the gap between the first and second detector modules, as shown in **Fig. 1**.

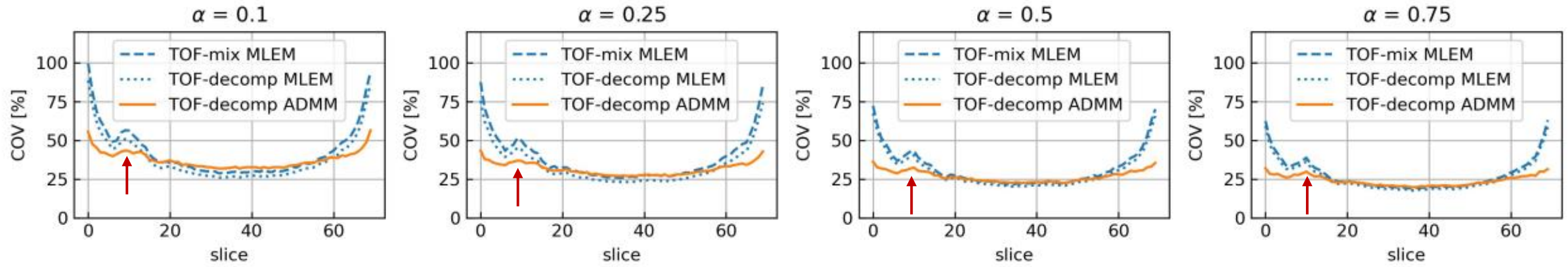

**Fig. 8.** Slice-wise COV obtained from IQ simulation data, evaluated at the iterations where $\mathrm{CRC_{ave}}$ reaches 95% of its maximum value are averaged over 10 samples. The dip in the slice-wise COV indicated by the red arrows is caused by the gap between the first and second detector modules, as shown in **Fig. 1**.

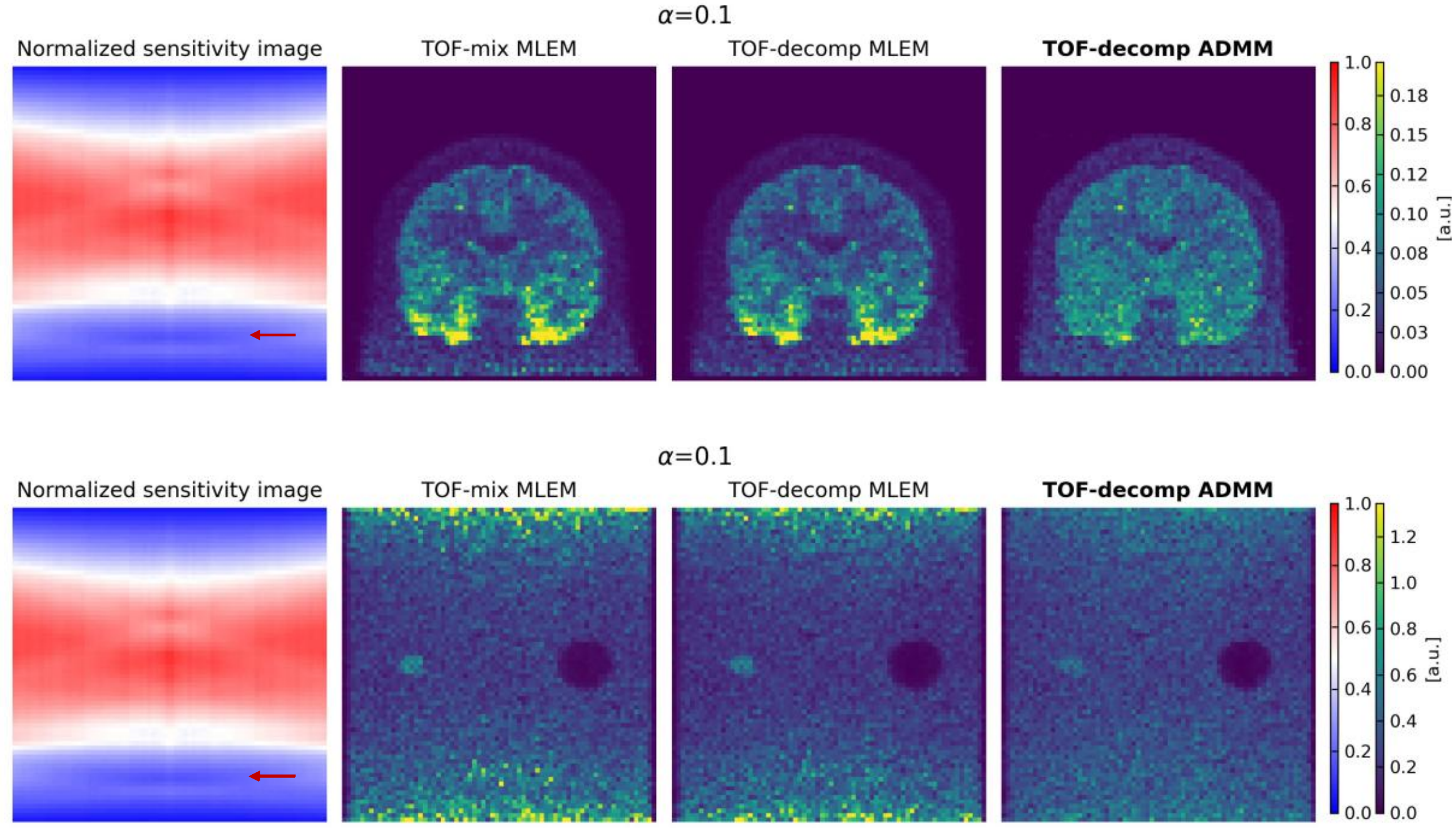

**Fig. 9.** Standard deviation images obtained from brain and IQ phantoms for $\alpha = 0.1$. Images are shown at the iterations that minimize RMSE for the brain phantom and at the iterations where $\mathrm{CRC_{ave}}$ reaches 95% of its maximum value for IQ phantom. The red arrows indicate the low-sensitivity region caused by the detector module gap.

Standard deviation images obtained from the brain and IQ phantom simulations for $\alpha = 0.1$ are shown in **Fig. 9**. The red arrows indicate the low-sensitivity region caused by the detector module gap. The TOF-decomp ADMM method effectively suppresses noise in these low-sensitivity regions, as well as near the axial edges of the field of view, highlighting its robustness against sensitivity nonuniformities. Meanwhile, a slight redistribution of noise toward more central regions (e.g., white-matter-like areas) can be observed. This behavior indicates differences in the spatial distribution and intensity dependence of noise between the MLEM and ADMM methods.

**Fig. 10** illustrates the convergence behavior and the relative progress of the fast- and slow-CTR log-likelihood terms obtained from the brain phantom simulations. Compared with the TOF-mix MLEM and TOF-decomp MLEM methods, the TOF-decomp ADMM method exhibits a more uniform balance in the progress of the two log-likelihood components. In contrast, the TOF-mix MLEM method shows a rapid decline in the progress of the fast-CTR log-likelihood, reflecting the difference between the mixed and decomposed objectives (Eqs. (3) and (6)). The TOF-decomp MLEM method, on the other hand, shows a relative preference toward the fast-CTR component.

Finally, **Fig. 11** compares TOF-decomp ADMM with an asymmetric update scheme ($M_1 = 2, M_2 = 1$) and a symmetric variant (TOF-decomp ADMM-sym, $M_1 = M_2 = 1$), obtained from brain simulation data. While the PSNR-TR and Bias$^2$-Var trade-off curves are similar between TOF-decomp ADMM and TOF-decomp ADMM-sym, the symmetric variant exhibits instability at very early iterations, as indicated by the red arrows.

## VI. Discussion

In this study, we proposed a multi-kernel TOF-PET image reconstruction method based on ADMM, referred to as TOF-decomp ADMM. The proposed TOF-decomp ADMM method aims to stabilize the convergence behavior between the

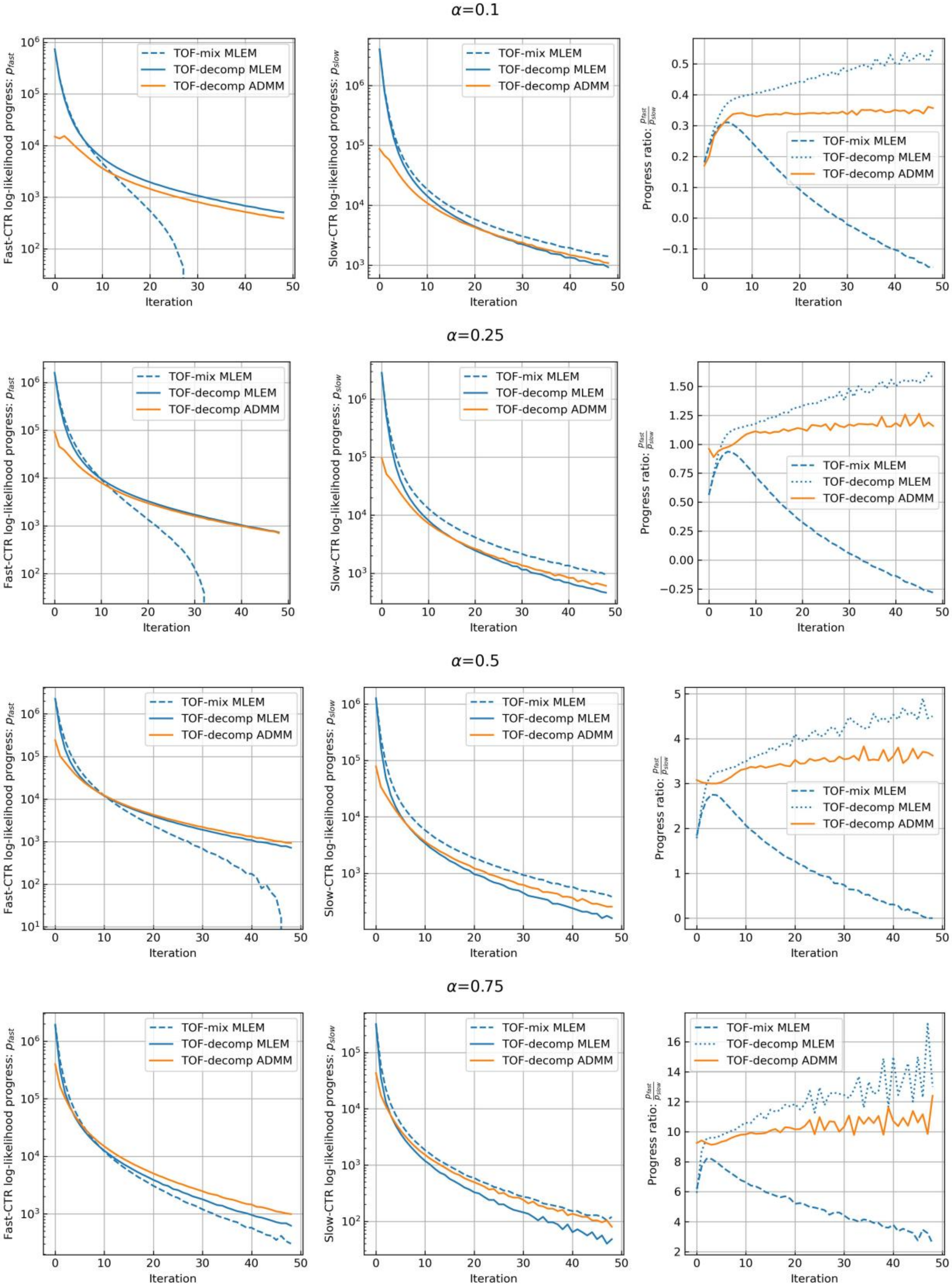

**Fig. 10.** Convergence behavior and relative progress of fast- and slow-CTR log-likelihood terms for different values of $\alpha$ obtained from brain simulation data.

fast- and slow-CTR log-likelihoods, thereby enabling more effective early stopping and achieving improved contrast–noise trade-offs.

Across various fast-CTR event ratios, the TOF-decomp ADMM method consistently achieved favorable contrast–noise and bias²–variance trade-off curves while maintaining a more uniform relative progress between fast- and slow-CTR log-likelihoods than TOF-mix MLEM and TOF-decomp

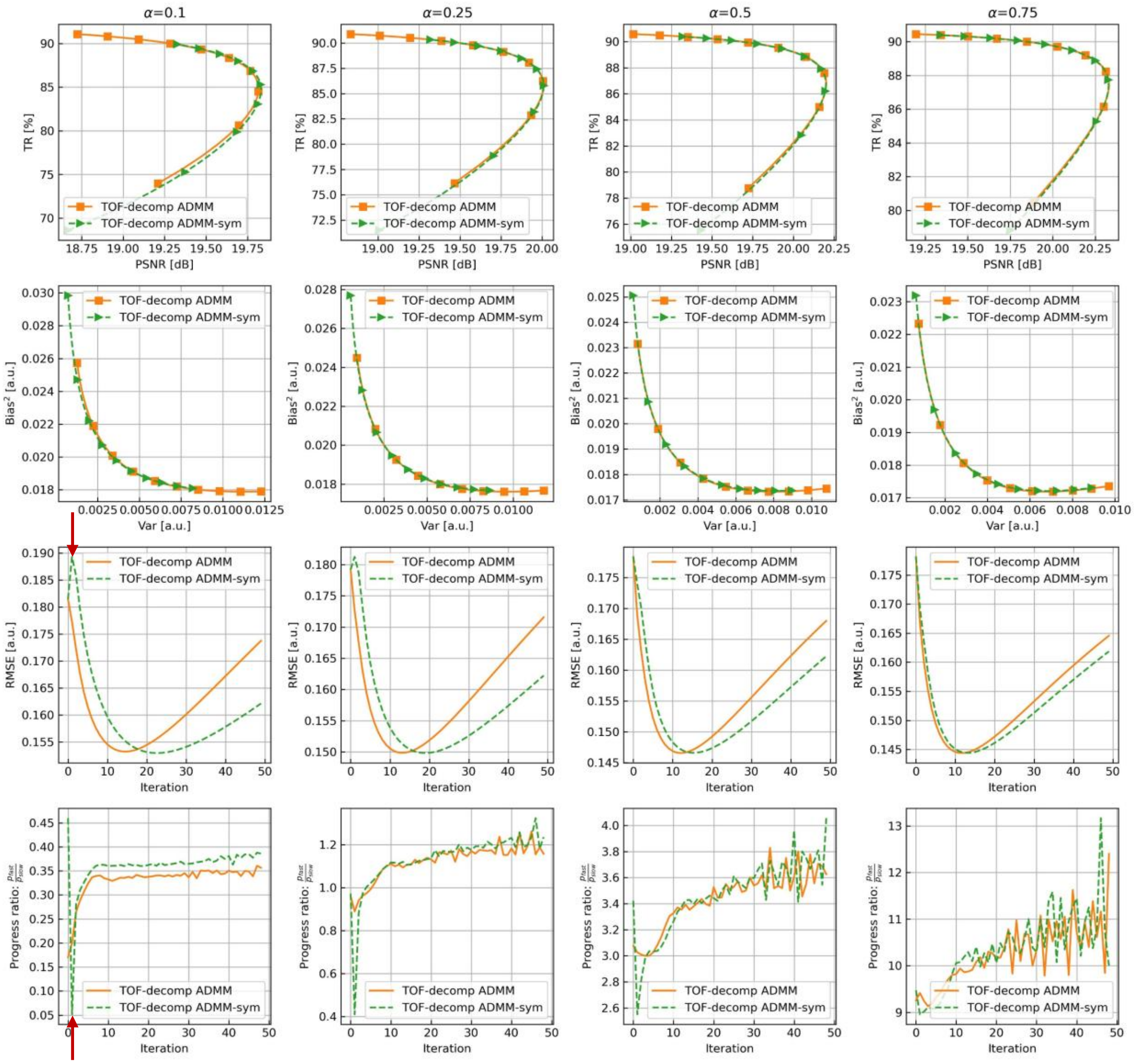

**Fig. 11.** Comparison between TOF-decomp ADMM with an asymmetric update scheme ($M_1 = 2, M_2 = 1$) and a symmetric variant (TOF-decomp ADMM-sym, $M_1 = M_2 = 1$), obtained from brain simulation data. The red arrows indicate the instability of TOF-decomp ADMM-sym at very early iterations.

MLEM methods (**Figs. 3, 6, and 10**). These observations suggest that explicitly balancing the progress of the fast- and slow-CTR components stabilizes the convergence behavior, thereby enabling effective stopping at iterations associated with superior contrast–noise trade-offs. Quantitative differences can be observed; however, the visual differences between the reconstruction methods remain relatively subtle (**Fig. 4**). This relatively subtle visual difference is consistent with the design of the proposed approach, which primarily balances the convergence behavior between fast- and slow-CTR components rather than introducing fundamentally new information. As a result, the improvements are mainly reflected in reconstruction stability and trade-off characteristics rather than in dramatic visual changes.

Previous studies have shown that improved CTR leads to enhanced cold-sphere contrast, particularly for larger spheres in phantom experiments [29], suggesting that the benefits of better CTR extend beyond the expected TOF-related signal-to-noise gain. In this work, we also observed an improvement in the contrast of larger cold spheres (**Fig. 5**), but this gain was achieved by separating events into fast- and slow-CTR components (TOF-decomp MLEM and TOF-decomp ADMM, versus TOF-mix MLEM) rather than through a uniform improvement in CTR. This indicates that event-wise CTR decomposition can influence contrast recovery beyond conventional SNR-based expectations.

The TOF-decomp ADMM method reduced noise and sample standard deviation in low-sensitivity regions caused by the detector module gap, as well as at the axial edges of the field of view (**Figs. 7–9**). Similar behavior has also been reported in previous studies in different contexts [30]. One possible explanation is that the penalty term in problem (9), which penalizes the $\ell_2$-norm of the difference between $\boldsymbol{x}$ and $\boldsymbol{z}$ induced by the consensus constraint ($\boldsymbol{x} = \boldsymbol{z}$) in problem (8), becomes larger in low-sensitivity regions due to increased statistical noise. Consequently, this penalty term acts to suppress noise in such regions. On the other hand, TOF-decomp ADMM method redistributes the noise toward more central regions, such as white matter in the brain phantom simulations (**Fig. 9**). Since MLEM methods based on Poisson log-likelihood function employ a multiplicative update scheme, the resulting noise is signal-dependent, with lower variance in low-intensity regions and higher variance in high-intensity regions [16]. In contrast, the ADMM framework combines multiplicative and additive update components through the introduction of the $\ell_2$-norm penalty term in the objective function. This transition from a purely multiplicative scheme to a combined multiplicative and additive formulation may reduce the intensity dependence of noise compared with the MLEM methods. This tendency could be adjusted by employing alternative penalty formulations, such as weighted $\ell_2$-norms or penalty terms based on Bregman divergence [31], which remains an interesting direction for future research. We note that, in the limit of convergence, the noise characteristics of TOF-decomp ADMM are expected to be consistent with those of TOF-decomp MLEM. However, difference in convergence behavior between MLEM and ADMM methods may allow control of noise characteristics prior to full convergence.

The symmetric variant of TOF-decomp ADMM, denoted as TOF-decomp ADMM-sym, exhibits instability at very early iterations, as shown in **Fig. 11**. This observation suggests that the asymmetric update plays an important role in stabilizing the reconstruction during the very early iterations, before the ADMM framework itself effectively stabilizes the optimization process. At this stage, it is difficult to attribute the observed improvements solely to either the update scheme or the ADMM framework itself. A systematic comparison with TOF-decomp MLEM using an equivalent asymmetric update schedule would be required to disentangle these effects and remains an important topic for future investigation. We additionally examined the RMSE curves as a function of the number of updates rather than the number of iterations (see Supplementary **Fig. S1**). For smaller fast-CTR event ratios (e.g., $\alpha = 0.1$), the RMSE curves of TOF-decomp ADMM and TOF-decomp ADMM-sym become more consistent when plotted against the number of updates, suggesting that the slow component is dominant and that the number of updates provides a more appropriate comparison axis in this case. In contrast, for larger fast-CTR ratios (e.g., $\alpha = 0.75$), the differences between the two methods become more pronounced in the update-based plot, suggesting that the fast component is dominant and that the number of iterations better reflects the convergence behavior. Therefore, the choice of the horizontal axis may depend on the fast-CTR event ratio.

In this study, we considered only two TOF kernels corresponding to fast and slow CTRs. However, in practice, more than two TOF kernels may exist, such as fast–slow and slow–fast CTR staggered events, which result from coincidences between asymmetric signals formed by faster- and slower-emitted photons populations. Such heterogeneous photons populations can occur in mixed Cherenkov-scintillation materials [1, 2]. In addition, heterostructured scintillators can give rise to even more complex CTR distributions due to variable energy sharing and photon emission dynamics [5]–[7]. Since ADMM is primarily designed to handle objective functions composed of the sum of two functions, extending the proposed framework to more than two TOF kernels would require alternative optimization schemes, such as Dykstra-like splitting [30] or stochastic primal–dual hybrid gradient methods [32], which are capable of handling the sum of multiple component functions. Extending the proposed framework to such multi-kernel settings therefore constitutes an important direction for future research adapted to novel detector concepts currently in development. In addition, kernel asymmetry arising from fast–slow or slow–fast events is expected to induce shifts in the effective TOF kernel peak position, which should be corrected for accurate image reconstruction.

In addition, the effect of misclassification between fast- and slow-CTR events remains an important topic for future investigation from an engineering perspective. Although classification of events into multiple TOF kernels based on rise-time discrimination has been proposed [1], more accurate separation may be achieved by restoring precise time stamps using single-photon response deconvolution techniques [33], [34]. Such approaches have the potential to improve the accuracy of decomposing mixed TOF-kernel contributions and to enable more reliable identification of asymmetric or mixed-origin events in future multi-kernel TOF-PET systems. Another prospect for multi-TOF frameworks might lie in deep learning approaches, which have already been applied to multi-TOF detectors, such as the improvement of CTR in BGO [35] and event classification in heterostructured scintillators [36]. At the reconstruction level, deep learning is gaining significant traction [37], and multi-TOF could be explored through approaches that better exploit the information carried by different CTR classes.

In this study, the fast and slow CTRs were fixed at 70 and 400 ps, respectively, following a previous multi-kernel study [10]. When the set of CTRs changes, the hyperparameters of the proposed algorithm, such as the number of sub-iterations and the step size, may need to be adjusted accordingly. In addition, the step size that yields favorable contrast–noise trade-offs can depend on the count level. While systematic tuning of these hyperparameters was not the primary focus of this study, developing strategies for their robust and adaptive selection represents an important topic for future work in practical multi-kernel TOF-PET image reconstruction.

Although $\boldsymbol{x}$ is used as the final estimate in this study, $\boldsymbol{x}$ and $\boldsymbol{z}$ can be interpreted as complementary reconstructions corresponding to fast- and slow-CTR components. Their combination such as $(\boldsymbol{x}+\boldsymbol{z})/2$ may offer further improvements. Therefore, we also evaluated this and observed that the results obtained from $\boldsymbol{x}$, $\boldsymbol{z}$, and $(\boldsymbol{x}+\boldsymbol{z})/2$ become nearly identical after several iterations (see Supplementary **Fig. S2**), indicating strong consistency enforced by the ADMM framework. This observation further supports that the proposed approach primarily improves convergence behavior rather than altering the final solution.

## VII. Conclusion

In this study, we developed a multi-kernel TOF-PET image reconstruction method using ADMM, referred to as TOF-decomp ADMM, based on the idea that balancing the convergence behaviors of the fast- and slow-CTR log-likelihoods enables stopping the iterations at points that achieve improved contrast–noise trade-offs.

We evaluated the TOF-decomp ADMM method using brain and IQ phantom simulations in terms of contrast–noise trade-offs and the relative progress between the fast- and slow-CTR log-likelihoods. The results showed that the TOF-decomp ADMM method provides more favorable contrast–noise trade-offs and more stable relative progress between the two log-likelihood components than the TOF-mix MLEM and TOF-decomp MLEM methods.

Overall, these findings indicate that multi-kernel TOF-PET is an important and promising research topic not only from the detector perspective but also from the algorithmic perspective.

## Acknowledgment

Francis Loignon-Houle is supported by Generalitat Valenciana through CIAPOS/2023/133 postdoctoral grant.

SUPPLEMENTARY MATERIALS

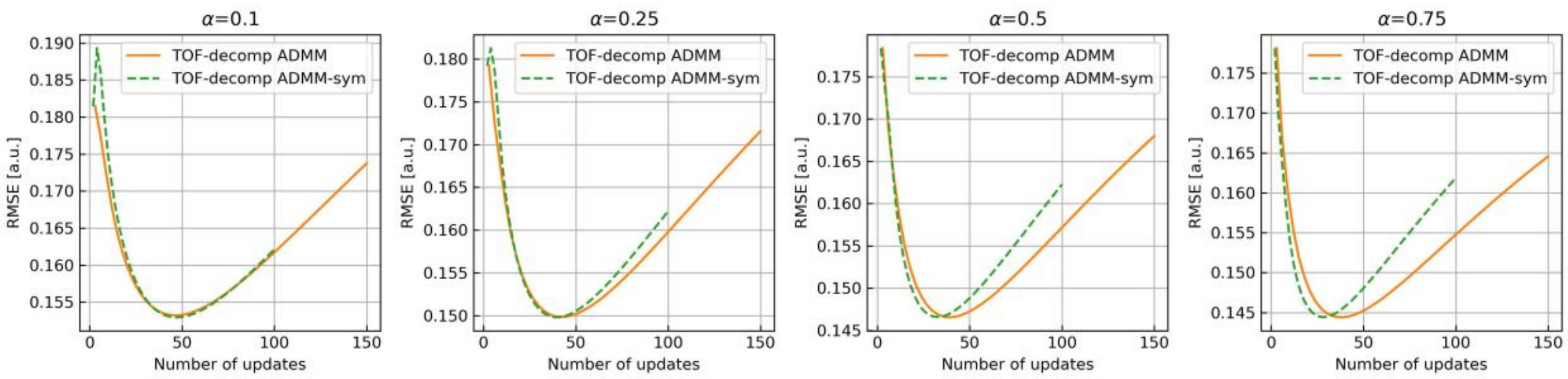

**Fig. S1** RMSE curves as a function of the number of updates, obtained from brain simulation data for the four different fractions of events with fast CTR.

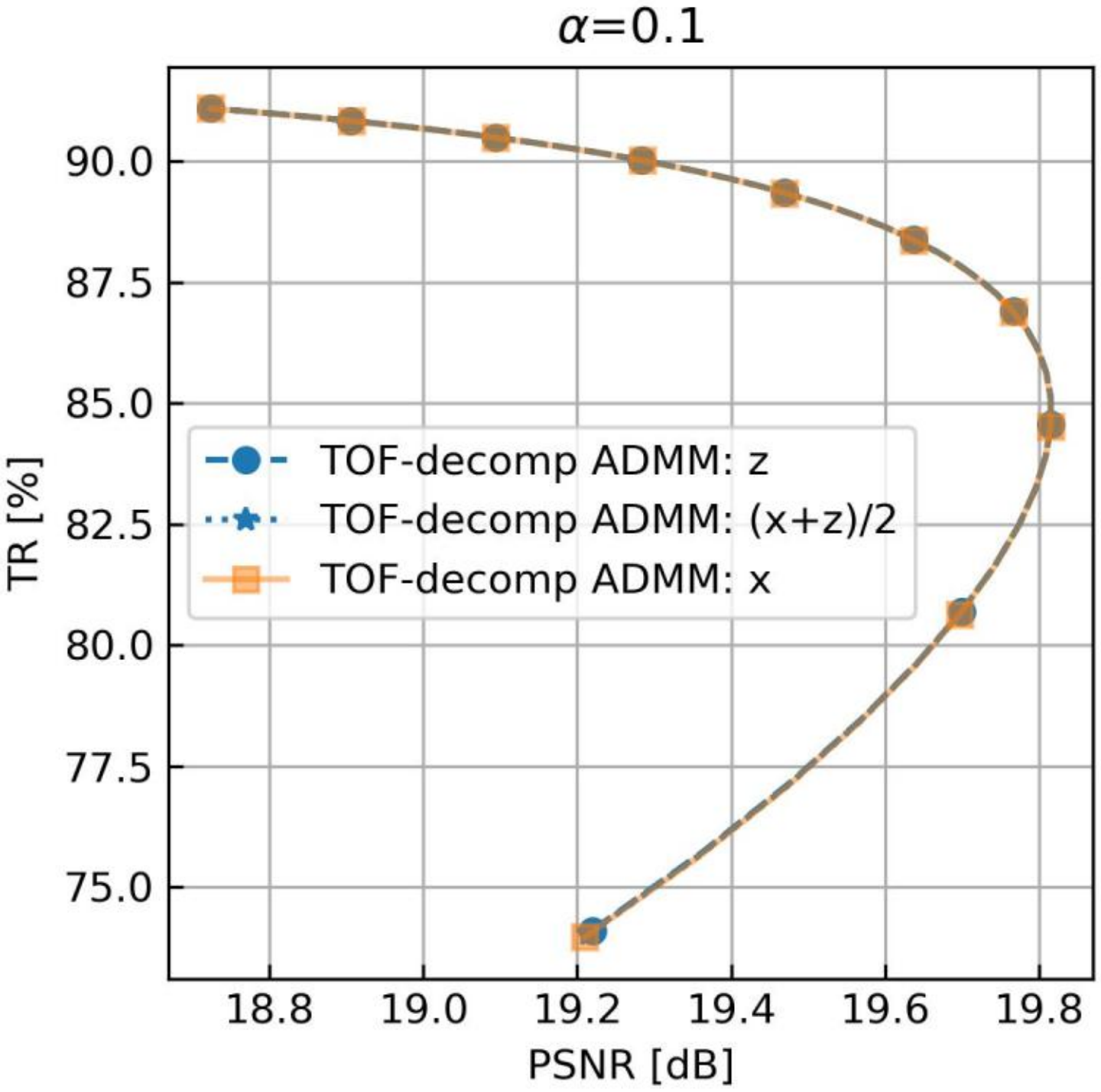

**Fig. S2** PSNR-TR curves of $\boldsymbol{x}$, $\boldsymbol{z}$, and $(\boldsymbol{x}+\boldsymbol{z})/2$ in TOF-decomp ADMM, obtained from the brain simulation data at $\alpha = 0.1$.